\documentclass{iaus}
\usepackage{graphicx}

\title[The Li Overabundance of J37] 
{The Li Overabundance of J37: \break Diffusion or Accretion?}

\author[J.F. Ashwell \textit{et al}]   
{J.F. Ashwell$^1$, R.D. Jeffries$^1$, B. Smalley$^1$, C.P. Deliyannis$^2$, \break A. Steinhauer$^2$ \and J.R. King$^3$}

\affiliation{$^1$Astrophysics Group, School of Chemistry and Physics, Keele University, \break Staffordshire, ST5 5BG, UK. email: jfa@astro.keele.ac.uk\\[\affilskip]

$^2$Department of Astronomy, Indiana University, Bloomington, IN47405-7105, USA\\[\affilskip]

$^3$Department of Physics and Astronomy, Clemson University, \break 118 Kinard Laboratory of Physics, Clemson, SC29634-0978, USA}

\pubyear{2004}
\volume{224}  
\pagerange{1-5}
\date{?? and in revised form ??}
\jname{The A-Star Puzzle}
\editors{J. Zverko, W.W. Weiss, J. Ziznovsky \& S.J. Adelman, eds.}
\begin{document}

\maketitle

\begin{abstract}
In September 2002 the discovery of a super Li-rich F-dwarf (J37) in NGC 6633, an iron poor
analogue of the better studied Hyades and Praecepe open clusters, was announced. This unique 
star was thought to be the smoking gun for the action of diffusion, models of which predict a 
narrow "Li-peak" at approximately the correct temperature. However, with more detailed studies 
into J37s abundance pattern this star provides firm evidence for the accretion of 
planetesimals or other material from the circumstellar environment of new born stars.

Thanks to the specific predictions made about the behaviour of Be abundances, (the most
striking of which being no Be in super-Li-rich dwarfs subject to diffusion) the opposing
diffusion/accretion predictions can be tested.

Initial modelling of the Be line indicates that J37 is as Be rich as it is Li rich; $\log N(Be)$ 
= 2.25 $\pm$ 0.25, and so is broadly consistent with an accretion-fuelled enhancement. However, 
that both Li and Be are enhanced by much more than the iron-peak elements (as determined in 
previous studies) suggests that diffusion also plays a role in increasing the abundances of Li 
and Be specifically.

Furthermore, a new data set from the UVES/UT2 combination has allowed the elemental abundance of 
Iron to be measured, and the set of preliminary stellar parameters determined; $\rm{T_{eff}}$ 
$\sim$ 7340 K, $\log g$ $\sim$ 4.1, microturbulence $\sim$ 4.3 $\rm{kms^{-1}}$, [Fe/H] $\sim$ 0.50.
This again provides distinct evidence for the effects of accretion in J37 and requires a new 
synthesis of the Be doublet.

\end{abstract}

\firstsection 

\section{Introduction}

The determination of Lithium and Beryllium photospheric abundances can reveal and improve the 
understanding of processes occurring in the stellar interior, with simultaneous measurements of 
both Li and Be being sensitive to mixing, diffusion and other poorly understood transport 
processes.

This work focuses specifically on the super Li-rich star, J37, discovered in September 2002 
which is a photometric, proper motion and radial velocity member of NGC6633 
(\cite[Deliyannis \etal\ 2002]{Deliyannis02}). The star itself is a late A or early F dwarf and 
as such should possess a very shallow convection zone as well as being a slow rotator for its 
spectral class at 29 $\pm$ 2 $\rm{kms^{-1}}$ (\cite[Deliyannis \etal\ 2002]{Deliyannis02}). This 
relatively thin mixing layer allows physical processes to produce observable effects that are 
mostly negligible in cooler dwarfs thus making J37 susceptible to abundance concentrations that 
might otherwise be diluted by turbulence, convection and meridional circulation in the atmosphere.

\cite[Deliyannis \etal\ (2002)]{Deliyannis02} suggested that the high Li abundance could be 
explained by the theoretical diffusion models of \cite[Richer \& Michaud (1993)]{Richer93}, as 
when internal mixing is insufficient to inhibit it, diffusion offers a viable mechanism for the 
Li overabundance. These diffusion models specifically predict an overabundance of Li in the very 
narrow $\rm{T_{eff}}$ range 6900-7100 K. In this $\rm{T_{eff}}$ range there is a substantial region 
below the extremely thin (model) surface convection zone where Li atoms retain one electron, and 
below this region are completely ionised and diffuse downwards (via gravitational settling and 
thermal diffusion) relative to hydrogen. By contrast, the electron retaining Li is radiatively 
accelerated upwards thus enriching the SCZ. It should be noted that subsequent improvements to 
these models (\cite[Richer \etal\ 2000]{Richer00}) results in them no longer showing a Li peak, 
but the reasons for this are unclear.

An alternative hypothesis comes in the form of accretion, with direct evidence for this process 
coming from either a correlation between metallicity and mass that could be explained 
in terms of decreasing CZ mass onto which the accreted material is mixed, or a correlation 
between the abundance enhancements of various elements and either the abundances found in 
planetesimals or the condensation temperature of each element (e.g. \cite[Gonzalez \etal\ 2001]{Gonzalez01}). It was \cite[Laws \& Gonzalez (2003)]{Laws03} who drew on accretion as the cause of J37's high Li 
abundance and concluded that, though neither diffusion or accretion of either type is in complete 
agreement with J37's abundance pattern the elements in the CZ of J37 were set by the composition 
of a) the primordial gas from the star's formation, b) subsequent accretion of both depleted 
circumstellar gas and planetesimal material, and c) evolution of its atmosphere through internal processes.

Despite this ambiguity one can, in principal, distinguish between stars exhibiting the effects 
of diffusion or accretion by looking simultaneously at other elements, such as Be, that burn at 
different temperatures or are stable in the stellar core. The Li/Be ratio is unlikely to be 
markedly changed by accretion processes, but because these are the lightest metals, they can 
behave in extreme and usually completely contrary ways, when subject to the effects of 
diffusion (e.g. \cite[Turcotte \etal\ 1998]{Turcotte98}). Thanks to the specific predictions made 
about the behaviour of Be abundances in the models of \cite[Richer \etal\ (2000)]{Richer00}, the 
most striking of which being the excessive underabundance of Be in super-Li-rich dwarfs, the opposing diffusion/accretion predictions can be tested.

\section{Observations \& Data Reduction}\label{sec:obs}

High resolution spectroscopy was obtained for J37 and 9 other stars in NGC 6633 with the 
UV-Visual Echelle Spectrograph (UVES) mounted on the UT2 8.2m VLT telescope at Cerro Paranal, 
Chile, on the night of 2003 June 12 -- 13. The standard DIC1/346/580 template with a 1.2 arcsec 
slit was used with 2 $\times$ 2 and 1 $\times$ 1 binning on the blue and red arms of the 
instrument respectively. This provides the wavelength range 3050 -- 6820\AA \ (orders 152 -- 90) 
at a resolution of 35\,000, with an exposure time of 3000.

The blue arm of UVES was used to measure the Be abundance while the red arm provided very high 
signal spectra allowing great improvements in the stellar atmospheric parameters (particularly 
$\rm{T_{eff}}$ and microturbulence), by detailed consideration of many (unblended) lines of 
neutral/ionised Fe and fitting the Balmer line profiles.

Due to problems with UVES's automatic data reduction pipeline all the data reduction and 
extraction was carried out using the LINUX computing facilities at Keele University and the 
ECHOMOP data reduction package. CCD frames were debiased, flat-fielded and sky-subtracted 
before the extraction of individual orders from the images. These extracted spectra were then 
wavelength calibrated using Thorium-Argon arc spectra taken on the same night.

\section{Preliminary Results \& Conclusions}\label{sec:pres}

\begin{figure}
 \centerline{
 \scalebox{1.2}{%
 \includegraphics{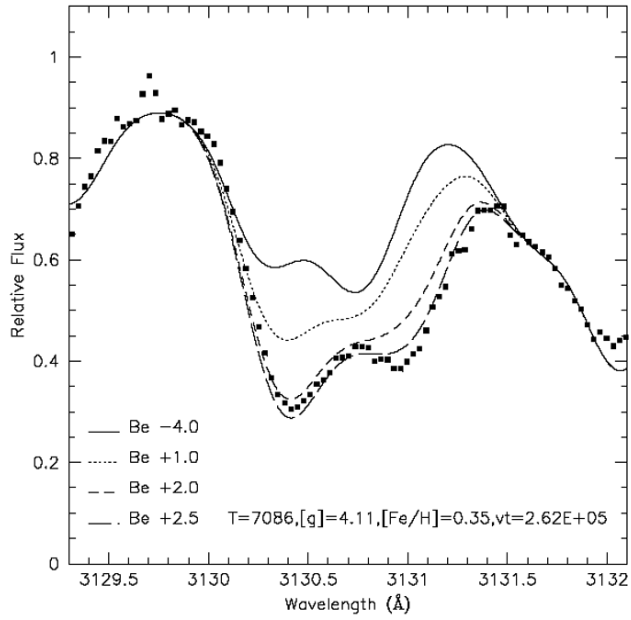}%
}
}
  \caption{An initial spectra synthesis of the region around the Be {\sc{ii}} doublet (3130.4 \AA \ 
  and 3131.1 \AA) in J37. The dotted trace shows the original data and the four syntheses show varying 
Be abundances, from $\log N(Be)$ = -4.0 dex (negligible Be) to $\log N(Be)$ = +2.5 dex (super solar). 
The currently favoured solution is log N(Be) = 2.25 $\pm$ 0.25 dex suggesting J37 is as Be rich 
as it is Li rich.}\label{fig:Be}
\end{figure}

Initial synthesis of the Be line using the parameters of \cite[Deliyannis \etal\ (2002)]{Deliyannis02} 
appears to favour the accretion hypothesis As mentioned in the introduction if diffusion is the 
correct solution for the Li overabundance there should be a very low Be abundance. An initial spectral 
synthesis of the Be~{\sc{ii}} doublet (3130.4 \AA \ and 3131.1 \AA) region is shown in 
figure~\ref{fig:Be}, and the favoured result is currently $\log N(Be)$ = +2.25 $\pm$ 0.25 dex suggesting 
the Be overabundance is almost as great at the Li overabundance. However, though the studies of iron peak elements carried out by \cite[Laws \& Gonzalez (2003)]{Laws03} 
and \cite[Deliyannis \etal\ (2002)]{Deliyannis02} show overabundances in these elements it is not to 
such an extent as Li and Be suggesting diffusion may also be playing a role in increasing the abundance 
of these particular elements.

The determination of a new set of atmospheric parameters started with the measurement of 44 unblended 
Fe {\sc{i}} and Fe {\sc{ii}} EWs between 4900 and 6800\AA \ using the gf-values held the {\sc{uclsyn}} 
code (\cite[Smith 1992]{Smith92}). EWs were measured by direct integration below a continuum that 
was estimated by fitting a low-order polynomial to line free regions of the surrounding spectrum. 
Initially working from the parameter set of \cite[Deliyannis \etal\ (2002)]{Deliyannis02} this line 
list was used to determine [Fe/H] via calculation of the mean abundance. The microturbulent velocity 
was then fixed by eliminating trends in Fe {\sc{i}} abundance with expected equivalent width following 
the determination techniques of \cite[Magain (1984)]{Magain84}. These quantities were then use in the 
determination of $\rm{T_{eff}}$ and $\log g$ values by balancing the mean Fe {\sc{i}} and Fe {\sc{ii}} 
abundances using the {\sc{uclsyn}} code. Selections of parameter sets from along locus were then taken 
and these new sets synthesised, with new [Fe/H] and microturbulence values calculated. This process was 
iterated (focused on a constant $\log g$ value) until the parameters settled. With several possible 
combinations of parameter sets possible; and in theory infinite numbers in between, a plot of $\rm{T_{eff}}$ 
against $\log g$ was produced.

\begin{figure}
 \centerline{
 \scalebox{0.8}{%
 \includegraphics{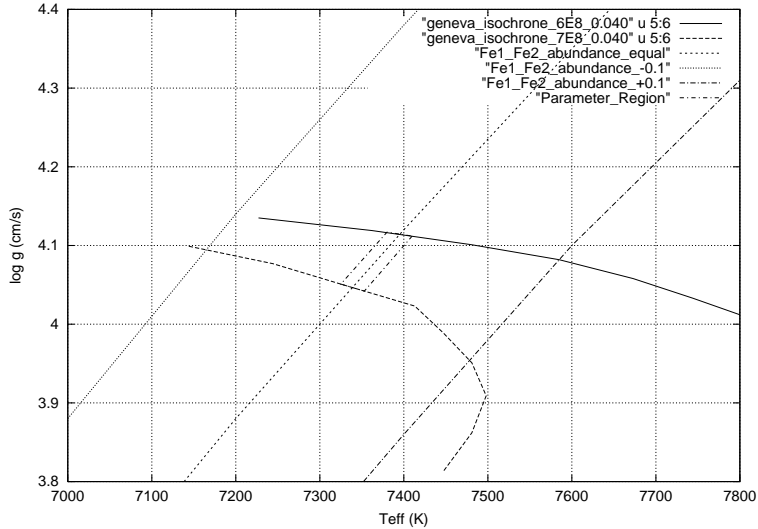}%
}
}
  \caption{Determination of J37 initial parameters via balancing the Fe {\sc{i}} and Fe {\sc{ii}} 
  abundances and limiting the balance to within $\pm$ 0.1 dex. These possible combinations of parameters 
  are limited via 600 My and 700 My Geneva stellar evolution isochrones. Finally, the upper limit of 
  the region of 1$\sigma$ variation in the gradient of excitation potential vs abundances of the Fe lines 
  from zero was calculated, all of which limits the initial parameter set to the boxed region.} \label{fig:parameters}
\end{figure}

The next stage, requiring the possible parameter sets to be limited, was to minimise the 
gradient via a least squares fit of excitation potential vs abundance with the Fe line list. 
Any gradient which fell within 1$\sigma$ of a zero gradient was conservatively deemed a 
reasonable threshold, and thus restricted the available parameter sets. It should be noted at 
this stage that none of the techniques utilised so far incorporates stellar evolution theory, 
and to this end the Geneva stellar evolution isochrones (\cite[Schaerer \etal\ 1993]{Schaerer93}) are 
incorporated at this stage. The [Fe/H] for J37 is clearly super solar and thus Z=0.040 isochrones were 
chosen (the highest metallicity available), with no convective overshoot and standard mass loss at ages 
600-700 My. The resulting parameter range for J37 is marked in figure~\ref{fig:parameters} 
as dictated by these three analysis techniques. From this first analysis the parameters are 
believed to be, $\rm{T_{eff}}$ $\sim$ 7340 K, $\log g$ $\sim$ 4.1, microturbulence $\sim$ 
4.3~$\rm{kms^{-1}}$ and [Fe/H] $\sim$ 0.50.

These initial parameters agree well with the results of \cite[Laws \& Gonzalez (2003)]{Laws03} and 
moves J37 off the predicted Li peak as predicted by \cite[Richer \& Michaud (1993)]{Richer93}. Therefore, 
this also provides evidence for the accretion process as the cause of J37s high Li abundance.

\end{document}